\documentclass{article}
\usepackage{spconf,amsmath,graphicx}
\usepackage{enumitem}
\usepackage{cite}
\usepackage{xcolor}
\usepackage{booktabs}
\usepackage{threeparttable}
\usepackage{amsfonts,bm}
\usepackage{mathtools}
\usepackage{pifont}% http://ctan.org/pkg/pifont
\usepackage{multirow}
\usepackage{algorithm}
\usepackage{algorithmic}
\usepackage{soul}

\usepackage{esvect}
\usepackage[bookmarks=true,breaklinks=true,colorlinks,linkcolor=blue,citecolor=blue,urlcolor=blue]{hyperref}
\usepackage{comment}
\usepackage{xspace}
\title{Quantum Neural Network Extraction Attack via Split Co-Teaching \vspace{-0.1in}
}

\name{Zhenxiao Fu\qquad Fan Chen\vspace{-12pt}}
\address{Department of Intelligent Systems Engineering, Indiana University, Bloomington, IN, USA\vspace{-6pt}}

\begin{document}
\maketitle

\begin{abstract}
Quantum Neural Networks (QNNs), now offered as QNN-as-a-Service (QNNaaS), have become key targets for model extraction attacks. Existing methods use ensemble learning to train substitute QNNs, but our analysis reveals significant limitations in real-world environments, where noise and cost constraints undermine their effectiveness.
In this work, we introduce a novel attack, \textit{split co-teaching}, which uses label variations to \textit{split} queried data by noise sensitivity and employs \textit{co-teaching} schemes to enhance extraction accuracy. The experimental results show that our approach outperforms classical extraction attacks by 6.5\%$\sim$9.5\% and existing QNN extraction methods by 0.1\%$\sim$3.7\% across various tasks.
\end{abstract}


\vspace{-0.1in}
\section{Introduction}
\label{sec:intro}
\vspace{-0.1in}

\begin{figure}[t!]
\centering
\includegraphics[width=\linewidth]{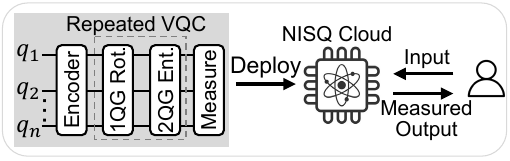}
\vspace{-0.35in}
\caption{The QNN-as-a-Service on a NISQ cloud server.}
\vspace{-0.2in}
\label{f:background_qnn}
\end{figure}

\textbf{Motivation}.
Quantum Neural Networks (QNNs) are powerful tools for complex problem-solving~\cite{schuld2015introduction, bharti2022noisy}, but their development requires specialized expertise and costly data, making them valuable intellectual property (IP) now offered as QNN-as-a-Service (QNNaaS)~\cite{pennylane, Google, QAI}, as illustrated in Figure~\ref{f:background_qnn}. This value has attracted adversaries who seek to steal QNNs from noisy intermediate-scale quantum (NISQ) cloud via model extraction~\cite{tramer2016stealing, Jacson:IJCNN2018, yu2020cloudleak, Fu:IJCNN2024}. 
Among these, QuantumLeak~\cite{Fu:IJCNN2024} stands out as the state-of-the-art (SOTA), utilizing ensemble learning to reduce noise in queried labels from QNNaaS, enabling the training of an accurate substitute QNN.
However, our experiments reveal that QuantumLeak faces substantial challenges in real-world environments. We increased the query rounds (e.g., from 3~\cite{Fu:IJCNN2024} to 5) to address varying quantum noise and reduced the number of queried data points (e.g., from 6000~\cite{Fu:IJCNN2024} to 3000) to minimize costs (e.g., \$1.6 per second for accessing IBM quantum computers~\cite{QIBM}) and maintain stealth. These adjustments led to a 7.49\%$\sim$9.85\% accuracy drop, rendering the QuantumLeak attack ineffective. % Consequently, we are exploring new QNN extraction attack strategies better suited to real-world conditions.

\noindent
\textbf{Contributions}.
This work makes the following contributions:
\begin{itemize}[leftmargin=*, topsep=-4pt, partopsep=-4pt, itemsep=-4pt]
\item % In Section~\ref{sec:prelim},
We identified a critical limitation in the SOTA QuantumLeak~\cite{Fu:IJCNN2024} attack, which employs ensemble learning to re-weight noisy labels. Under NISQ conditions, this method fails as excessive noise misguides the substitute QNN, necessitating pre-filtering to distinguish clean data from noise-prone samples.
Inspired by the success of the classical co-teaching strategy~\cite{han2018coteaching, jiang2018mentornet}, which employs multiple neural networks to classify and clean noisy data, we adapted co-teaching to the quantum domain. 
% However, its effectiveness is limited, as the loss-based classification of noisy and clean data, essential to co-teaching, does not effectively translate to quantum settings.

\item % In Section~\ref{sec:design},
We propose a quantum co-teaching framework, \textit{split co-teaching}, to overcome the limited effectiveness observed in our preliminary study when directly adapting classical co-teaching to the quantum setting.
This approach involves \textit{splitting} noise-robust data from noise-vulnerable data based on variations in queried labels obtained at different times, followed by \textit{co-teaching} QNNs with configurations tailored to the noise sensitivity of the data.

\item % In Section~\ref{sec:exp},
We implemented the framework on a NISQ device and compared the local substitute QNN's accuracy with QNNs trained via SOTA extraction attacks. 
Results show that our approach outperforms the classical extraction attack~\cite{yu2020cloudleak} by 6.5\%$\sim$9.5\% and the QNN extraction attack~\cite{Fu:IJCNN2024} by 0.1\%$\sim$3.7\% across various tasks.
\end{itemize}

%%%%%%%%%%%%%%%%%%%%%%%%%
%%%%%%%%%%%%%%%%%%%%%%%%%

%%%%%%%%%%%%%%%%%%%%%%%%%%%%%%%%%%%%%%%%%%%%
\begin{figure*}[t!]
%%%%%%%%%%%%%
\begin{minipage}{0.31\linewidth}
\vspace{0.15in}
\centering
\includegraphics[width=\linewidth]{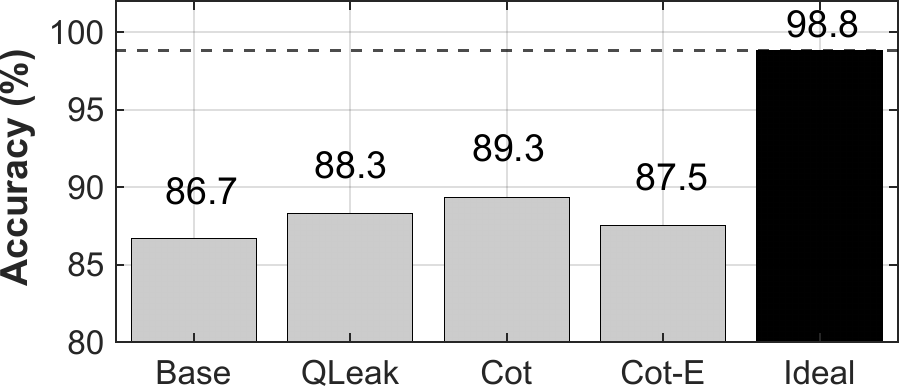}
\vspace{-0.33in}
\caption{Ineffectiveness of existing model extraction techniques.}
\label{f:baseline}
\end{minipage}
%%%%%%%%%%%%%
\begin{minipage}{0.69\linewidth}
\centering
\includegraphics[width=\linewidth]{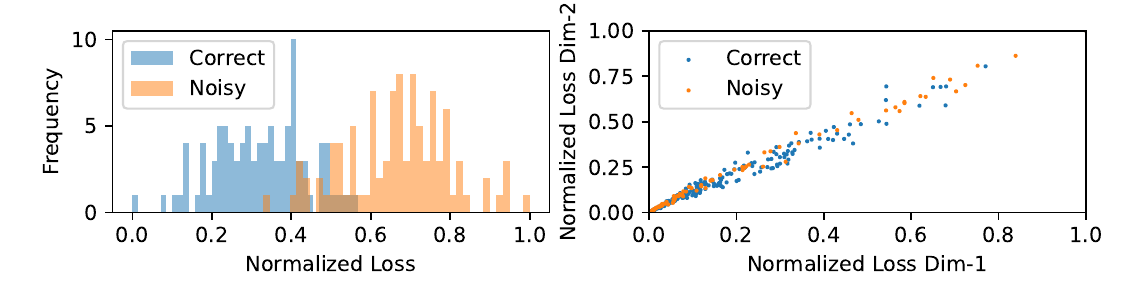}
\vspace{-0.4in}
\caption{Loss after one epoch for 300 noisy data points: (a) classical and (b) quantum.}
\label{f:err}
\end{minipage}
%%%%%%%%%%%%%
\vspace{-0.25in}
\end{figure*}  
%%%%%%%%%%%%%%%%%%%%%%%%%%%%%%%%%%%%%%%%%%%%

\vspace{-0.1in}
\section{Background}
\label{sec:back}
\vspace{-0.1in}

\textbf{QNN Basics}.
Quantum neural networks are a representative NISQ algorithm capable of operating on current noisy quantum computers. Central to a QNN are Variational Quantum Circuits (VQCs)~\cite{chen2020variational, Chu:ISLPED2022, chu2023qtrojan}, which are parameterized circuit ansatz typically implemented with one-qubit gates (i.e., 1QG) for rotation and two-qubit gates (i.e., 2QG) for entanglement. 
As shown in the shaded block of Figure~\ref{f:background_qnn}, a standard QNN comprises three key components: 
(1) a data encoder that embeds data into quantum states; 
(2) a multi-layer, trainable VQC circuit; and 
(3) a measurement layer that projects quantum states into probabilistic vectors. 
QNN generate the final prediction by applying an activation function (e.g., \texttt{softmax}) to the raw probability output vector.

\textbf{Quantum Noises}.
NISQ hardware suffers from various types of noise~\cite{Huo:NJP2017, brownnutt2015ion, Gustavsson:SCI2016} due to environmental interactions, imperfect controls, cross-talk, and other factors. These noise sources introduce quantum errors, such as decoherence, which leads to information loss; gate errors, causing operational inaccuracies; and readout errors, resulting in incorrect measurements.
Moreover, quantum noise in NISQ platforms is not static; it can fluctuate spatially and temporally.
For example, in trapped-ion systems~\cite{brownnutt2015ion}, fluctuations can arise from instabilities in laser and voltage control, while superconducting qubits~\cite{Gustavsson:SCI2016} are affected by variations in unpaired electron populations.
Table~\ref{tab:quantum_noise_change} illustrates this variability, showing that gate errors on a NISQ computer, measured at different times on the same day, exhibited a 31.2\% variation (i.e., $\Delta$) for one-qubit gates and a 10.3\% variation for two-qubit gates.

\begin{table}[h!]
\footnotesize
\vspace{-0.2in}
\caption{Gate error rates on \texttt{IBM\_Quebec} (6/30/2024).}
\label{tab:quantum_noise_change}
\centering
\setlength{\tabcolsep}{4pt}
\vspace{0.02in}
\begin{tabular}{|c|c|c|c|c|c|}\hline
\textbf{Time}       & 6:00     & 12:00    & 18:00     & 24:00 & \textbf{$\Delta$} \\\hline\hline
\multicolumn{1}{|l|}{\textbf{1QG (qubit 2)}} & 1.973e-4 & 2.343e-4 & 1.786e-4  & 1.942e-4 & \textbf{31.2\%}\\\hline
\multicolumn{1}{|l|}{\textbf{2QG (qubit 2/3)}} & 4.561e-3 & 5.033e-3 & 4.985e-3  & 4.582e-3 & \textbf{10.3\%} \\
\hline
\end{tabular}
\vspace{-0.1in}
\end{table}

\textbf{Co-Teaching}.
Classical co-teaching~\cite{han2018coteaching, jiang2018mentornet} effectively trains deep learning models with noisy labels. It leverages the tendency of neural networks to memorize clean labels before noisy ones, as indicated by lower loss values for clean data in early epochs. In co-teaching, two neural networks are trained simultaneously: each selects data with potentially clean labels (i.e., lower loss values) from a mini-batch, shares these selections with its peer, and backpropagates using the data chosen by the other. This process significantly enhances model robustness and demonstrates superior performance.

% \textbf{QNN-as-a-Service}.
% Due to the high costs and complexities of NISQ hardware and algorithm design, individual users often face significant access barriers to QNNs. Consequently, QNN-as-a-Service (QNNaaS)~\cite{pennylane, Google, QAI} has emerged as a vital cloud service. As illustrated in Figure~\ref{f:background_qnn}, users submit queries with their local data to the cloud server, which executes the QNN on its NISQ devices, measures the outputs, and returns classical probability results for further processing, ultimately producing a classical prediction. This model restricts user access to essential information, such as the QNN architecture, the training datasets used by the provider, and other hardware specifics like error rates and power consumption.

\vspace{-0.1in}
\section{Preliminary Study and Motivation}
\label{sec:prelim}
\vspace{-0.07in}

\subsection{Preliminary Study}
\vspace{-0.06in}

\textbf{Ineffectiveness of Existing Techniques}.
We trained a victim QNN using a recent model~\cite{Chu:ISLPED2022} on the MNIST dataset,
achieving 98.8\% accuracy in simulation.
This accuracy, denoted as \textit{Ideal}, represents the upper bound a substitute QNN could achieve.
To reduce the amount of queried data and increase the number of query rounds, we selected 3,000 data samples (compared to 6,000 in~\cite{Fu:IJCNN2024}) and queried the victim QNN 5 times (compared to 3 times in~\cite{Fu:IJCNN2024}) over a 24-hour period to obtain labels. We implemented model extraction attacks using classical CloudLeak~\cite{yu2020cloudleak}, QuantumLeak~\cite{Fu:IJCNN2024}, classical co-teaching~\cite{han2018coteaching}, and their combination.
All QNNs were deployed on \texttt{IBM\_Quebec}. For detailed information on datasets, NISQ configuration, and comparison schemes, please refer to Section~\ref{sec:exp}.
As shown in Figure~\ref{f:baseline}, no existing techniques, whether used individually or in combination, can construct a local substitute QNN that approaches the performance of the victim QNN. Furthermore, these experimental results reveal several previously undiscovered key findings:
\begin{itemize}[leftmargin=*, topsep=-2pt, partopsep=-2pt, itemsep=-2pt]
\item \textit{Observation-1}: Considering the noise fluctuation of NISQ devices, QuantumLeak~\cite{Fu:IJCNN2024} only achieves a $<$2.5\% accuracy improvement over the classical CloudLeak~\cite{yu2020cloudleak} attack, rather than the reported 4.99\%$\sim$7.35\% improvement.
\item \textit{Observation-2}: The classical co-teaching~\cite{han2018coteaching} is effective, showing a 1.8\% accuracy improvement over QuantumLeak~\cite{Fu:IJCNN2024} when na\"ively adopted without optimization.
\item \textit{Observation-3}: Integrating the ensemble-learning approach from QuantumLeak~\cite{Fu:IJCNN2024} into co-teaching~\cite{han2018coteaching} diminishes performance rather than enhancing it.
\end{itemize}

\vspace{2pt}
\noindent
\textbf{Classical vs. Quantum Noisy Data}.
Following~\cite{arpit2017closer}, we compare loss distributions after one epoch to assess the impact of noisy labels on classical and quantum neural networks.
Specifically, we analyze the classical CloudLeak~\cite{yu2020cloudleak} attack for classical model extraction using cross-entropy loss, and QuantumLeak~\cite{Fu:IJCNN2024} for QNN extraction, which uses a 2-dimensional Huber loss.
As shown in Figure~\ref{f:err}, classical models tend to learn clean patterns in the initial epochs, allowing small-loss instances to be easily filtered out as clean data. This aligns with the conclusions in~\cite{arpit2017closer} and is effectively leveraged in co-teaching~\cite{han2018coteaching}. In contrast, this correlation does not exist in QNNs, leading to our fourth observation:
\begin{itemize}[leftmargin=*, topsep=-4pt, partopsep=-4pt, itemsep=-4pt]
\item \textit{Observation-4}: The clean and noisy data are nearly evenly distributed across specific loss values, blurring the distinction between the two. This makes it difficult to effectively filter out clean data based on loss, a method that is typically successful in classical models~\cite{arpit2017closer, han2018coteaching}. % Furthermore, most loss values are concentrated at lower levels.
\end{itemize}

\vspace{-0.1in}
\subsection{Motivation}
\vspace{-0.1in}
\textit{Observation-1}, \textit{2} and \textit{4} are interconnected and mutually informative.
Specifically, the quantum noisy fluctuations render QuantumLeak~\cite{Fu:IJCNN2024}, the SOTA QNN extraction attack, ineffective, underscoring the importance of training approaches that account for noisy labels. However, the unique interaction between quantum noise and QNNs undermines the theoretical foundation of co-teaching~\cite{han2018coteaching}, highlighting the need for a quantum-specific data partition scheme.
\textbf{This motivates us to optimize co-teaching by partitioning data based on noise sensitivity rather than the loss values used in classical methods.}. 
Additionally, for \textit{Observation-3}, we hypothesize that the multiple ensembles and bagging in QuantumLeak degrade data quality by increasing the proportion of incorrectly labeled instances, leading to reduced performance. This will be investigated further in future studies.

%%%%%%%%%%%%%%%%%%%%%%%%%%%%%%%%%%%%%%%%%%%%
\begin{figure*}[t!]
%%%%%%%%%%%%%
\begin{minipage}{0.40\linewidth}
\begin{center}
\includegraphics[width=\linewidth]{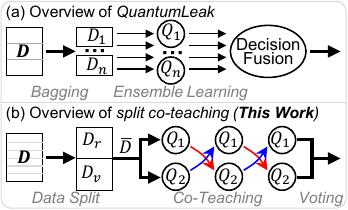}
\vspace{-0.32in}
\caption{Comparison of workflow between SOTA QuantumLeak~\cite{Fu:IJCNN2024} attack and this work.}
\label{f:overall}
\end{center}
\end{minipage}
\hspace{0.01in}
%%%%%%%%%%%%%
\begin{minipage}{0.60\linewidth}
\begin{center}
\includegraphics[width=\linewidth]{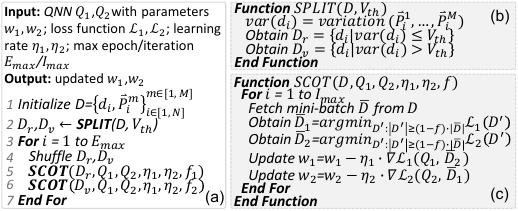}
\vspace{-0.32in}
\caption{Pseudocode for (a) proposed split co-teaching overview, utilizing algorithm (b) queried data split and (c) co-teaching method.}
\label{f:cot-alg}
\end{center}
\end{minipage}
\vspace{-0.25in}
\end{figure*}  
%%%%%%%%%%%%%%%%%%%%%%%%%%%%%%%%%%%%%%%%%%%%

\vspace{-0.18in}
\section{Split Co-Teaching}
\label{sec:design}
\vspace{-0.1in}
To address the limitations of existing QNN extraction attacks, which underestimate quantum noise fluctuations and overestimate the available number of queried data, we propose a new attack that incorporates practical NISQ features to train an accurate substitute QNN. As shown in the workflow comparison with the SOTA QuantumLeak~\cite{Fu:IJCNN2024} in Figure~\ref{f:overall}, 
our approach introduces two key improvements: (1) it captures quantum noise fluctuations by conducting more query rounds than QuantumLeak, and (2) instead of relying on an ensemble of QNNs with data bagging—an approach that incurs high design costs and has proven ineffective in our preliminary results—we split data based on quantum noise sensitivity and employ co-teaching to address training with noisy labels. 
% Detailed design and implementation are provided in the following sections.

\vspace{-0.15in}
\subsection{Queried Data Split}
\vspace{-0.05in}
Similar to the QuantumLeak attack~\cite{Fu:IJCNN2024}, we use publicly available data to query QNNaaS, with queries evenly distributed throughout the day to capture the impact of fluctuating noise.
Specifically, data is sent to the victim QNN in $M$ (e.g., $M$=5) rounds of queries, spaced at intervals of $24/M$ hours. The server executes the QNN on its NISQ devices, measures the outputs, and returns classical probability results.
As shown in Figure~\ref{f:cot-alg}(a), 
the resulting dataset is represented as $D$=\{$d_i$, $\vec{P}^{\,m}_{i}$\}, 
where $\vec{P}$ denotes the obtained raw probability output vectors,
$i\in$[1,$N$] represents the number of data samples, and 
$m\in$[1,$M$] corresponds to the query rounds.

The \texttt{SPLIT} function in Figure~\ref{f:cot-alg}(b) calculates the variation among the raw probability output vectors for each data sample $d_i$ across different query rounds, resulting in $var$($d_i$)$\in$[0,1].
A lower value indicates that $d_i$ is robust to NISQ noise, while a higher value indicates vulnerability. We set a predefined threshold $V_{th}$ to classify the dataset $D$ to a robust subset $D_r$ (with $var$($d_i$)$\leq$$V_{th}$)
and a vulnerable subset $D_v$ (with $var$($d_i$)$>$$V_{th}$).

\begin{figure*}[h]
\centering
\includegraphics[width=1\linewidth]{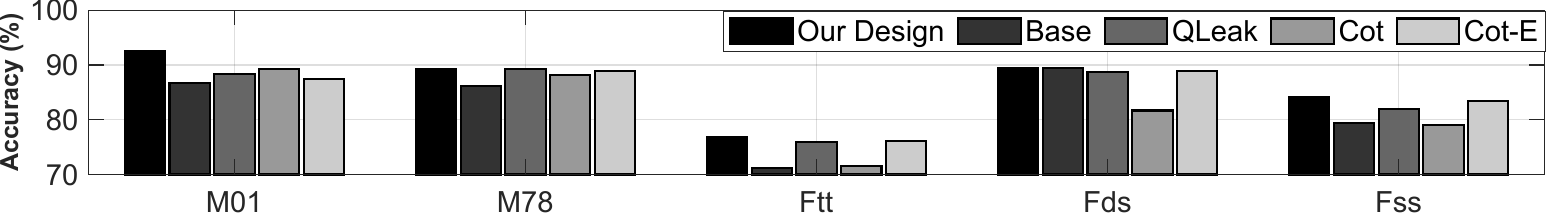}
\vspace{-0.35in}
\caption{Comparison of accuracy between this work and existing approaches across different tasks.}
\label{f:diftasks}
\vspace{-0.15in}
\end{figure*}
%%%%%%%%%%%%%%%%%%%%%%%%%%%%%%%%%%%%%%%%%%%%
\begin{figure*}[t!]
%%%%%%%%%%%%%
\begin{minipage}{0.32\linewidth}
\centering
\includegraphics[width=\linewidth]{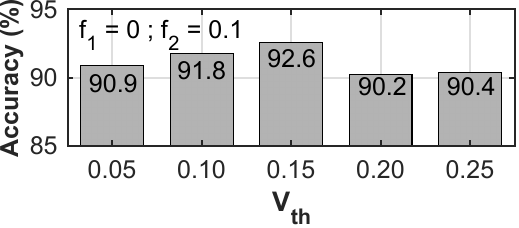}
\vspace{-0.35in}
\caption{Impact of $V_{th}$ on accuracy.}
\label{f:vth}
\end{minipage}
\hspace{2pt}
%%%%%%%%%%%%%
\begin{minipage}{0.32\linewidth}
\centering
\includegraphics[width=\linewidth]{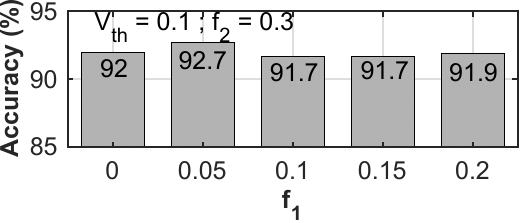}
\vspace{-0.35in}
\caption{Impact of $f_1$ on accuracy. }
\label{f:f1}
\end{minipage}
\hspace{2pt}
%%%%%%%%%%%%%
\begin{minipage}{0.32\linewidth}
\centering
\includegraphics[width=\linewidth]{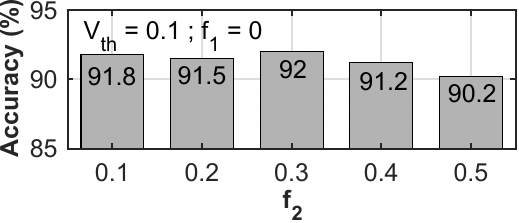}
\vspace{-0.35in}
\caption{Impact of $f_2$ on accuracy. }
\label{f:f2}
\end{minipage}
%%%%%%%%%%%%%
\vspace{-0.26in}
\end{figure*}  
%%%%%%%%%%%%%%%%%%%%%%%%%%%%%%%%%%%%%%%%%%%%

\vspace{-0.15in}
\subsection{QNN Co-Teaching}
\vspace{-0.05in}
The proposed approach involves training two QNNs simultaneously, where each QNN identifies and selects a subset of ``clean'' data from each mini-batch based on its confidence—specifically, instances that result in smaller loss values. 
This selected subset of data is then shared with the peer QNN, which uses it to update its parameters for the subsequent mini-batch. This collaborative process aims to enhance the training robustness of both QNNs by leveraging the strengths of each network. 
As illustrated in Figure~\ref{f:overall}(b), we refer to this method as split co-teaching. 
The overall algorithmic workflow is provided in Figure~\ref{f:cot-alg}(a) (lines 3-7), with the detailed implementation of the split co-teaching function (\texttt{SCOT}) outlined in Figure~\ref{f:cot-alg}(c).

In this approach, two QNNs, $Q_1$ and $Q_2$, with corresponding parameters $w_1$ and $w_2$, are trained simultaneously. 
During each training iteration, the mini-batch of data, denoted as $\overline{D}$, is processed by each QNN to identify and select a subset of instances that produce smaller losses, indicating higher confidence in the correctness of those instances. For example, $Q_1$
might select a subset $D_1$ as its most reliable data. This subset is then considered useful knowledge for training and is subsequently shared with the peer QNN ($Q_2$) to update its parameters.
The process of selecting these instances is regulated by a forget rate $f$ which determines the proportion of large-loss instances within the mini-batch that are excluded from training.
Furthermore, different forget rates, such as $f_1$ for $D_r$ and $f_2$ for $D_v$, are applied depending on the data's sensitivity to quantum noise. 
Specifically, a smaller forget rate $f_1$ is applied to $D_r$ to preserve its robustness, while a larger forget rate $f_2$ is used for $D_v$ to penalize its higher sensitivity to noise, thereby refining the training process for each QNN in a manner tailored to the characteristics of the data.

% we are effectively discarding a roughly equal number of noisy and clean data points, still resulting in an increased proportion of clean data. For instance, in the scenario of 300 samples depicted in Figure~\ref{f:err}, where there are 215 clean and 85 noisy samples (clean data proportion of 71.67\%), by simply discarding 11 clean and 11 noisy data points as their have the largest sums of Huber-loss, the clean data proportion rises to 73.38\%.

% Theoretically, this method could continuously increase the proportion of clean data by significantly discarding samples; however, the practical scenario differs. In model extraction attacks, the number of queries we can perform is extremely limited, and we still require a sufficient amount of data to train our local models, hence the proportion of data to be discarded (i.e., the Co-teaching forget rate) is also a parameter that needs balancing, which will be further studied in the following sections.

\vspace{-0.2in}
\subsection{NISQ Implementation}
\vspace{-0.08in}
The work focuses on proposing a new attack framework that addresses practical NISQ noise variations and reduced queried data in QNN model extraction attacks, rather than on fine-tuning the QNN itself. 
To ensure an apples-to-apples comparison, we adopt all implementation configurations from the QuantumLeak~\cite{Fu:IJCNN2024} attack, utilizing the same victim QNN circuit and local substitute QNN zoo, but with half the number of queried data samples.

% \section{Potential Defenses against VQC Stealing}
% \label{s:defenses}
% Through watermarking~\cite{Jia:Security2021, yang2024multi}, a VQC owner can establish ownership by inspecting VQCs suspected of being stolen. Watermarks involve overfitting the VQC to outlier input-output pairs that only the defender knows. Later, these watermarks can be used to assert ownership of the VQC. Outliers are created by inserting a unique trigger into the input, like a small square in a non-intrusive location of an input data sample. These specific inputs serve as watermarks. If the defender encounters a model that exhibits the rare and unexpected behavior encoded by the watermarks, she can reasonably conclude that this model is a stolen replica.

\vspace{-0.15in}
\section{Experiments and Results}
\label{sec:exp}

\vspace{-0.08in}
\subsection{Experimental Setup}
\vspace{-0.05in}

\textbf{Datasets}. 
Following recent QNNs~\cite{chen2020variational, Chu:ISLPED2022}, we used MNIST~\cite{mnist} and Fashion-MNIST~\cite{xiao2017fashion} datasets, down-sampling data to 1$\times$8 vectors using principal component analysis.
For each dataset, 3000 images were sampled to query the QNNaaS server five times daily, while 500 images were selected for validation and 1000 for testing. 
For MNIST, we perform classification of digits 0/1 (\texttt{M01}) and 7/8 (\texttt{M78}). For Fashion-MNIST, classification tasks included t-shirt/trouser (\texttt{Ftt}), dress/sneaker (\texttt{Fds}), and shirt/sneaker (\texttt{Fss}).

\textbf{Configuration and QNN Zoo}.
We trained a victim QNN for each task, implemented with four qubits. Each QNN architecture included one amplitude encoding layer, two repeated VQC layers, and one measurement layer. The VQC ansatz included an \texttt{RZ}-\texttt{RY}-\texttt{RZ} rotation layer and a 2-qubit \texttt{CRX} entanglement layer. We employed the Adam optimizer with a learning rate of 5e-3 and a weight decay of 1e-4, training with a batch size of 32 over 30 epochs.
We adopted the same local QNN zoo as used in QuantumLeak~\cite{Fu:IJCNN2024}, including the victim QNN, \cite{chen2020variational}, and \cite{Chu:ISLPED2022}.
All quantum circuit designs were synthesized using Pennylane~\cite{bergholm2018pennylane} and deployed to NISQ devices with Qiskit~\cite{Alexander:QST2020}. The circuits were executed and measured on the 127-qubit \texttt{IBM\_Quebec} computer.

\textbf{Schemes}. 
% previous work: 1-cloudleak (base); 2-Qleak; 3-Co-Teaching: 4-co-teaching + ensemble 
To compare our design with SOTA model extraction attacks, we established following four baselines:
\begin{itemize}[leftmargin=*, nosep, topsep=0pt, partopsep=0pt]
\item \textit{Base}: A local substitute QNN identical to the victim QNN. We applied the state-of-the-art classical technique, CloudLeak~\cite{yu2020cloudleak}, to extract the QNN using queried data.
\item \textit{QLeak}: We adopted the QNN architecture with the best performance as reported in~\cite{Fu:IJCNN2024} and trained it using the ensemble learning method detailed in the same study.
\item \textit{CoT}: A local substitute QNN selected from the QNN zoo, trained with classical co-teaching~\cite{han2018coteaching}, where two QNNs were simultaneously trained, each using data selected by its peer QNN for parameter updates.
\item \textit{CoT-E}: A combined technique of QuantumLeak~\cite{Fu:IJCNN2024} and co-teaching~\cite{han2018coteaching} where an ensemble of the best QuantumLeak substitute QNNs is trained with co-teaching.
\end{itemize}

\vspace{-0.15in}
\subsection{Results and Analysis}
\vspace{-0.05in}

Similar to~\cite{Fu:IJCNN2024}, our results show that a local QNN using the same VQC ansatz as the victim QNN achieves the best accuracy. The following results are based on this configuration.

\textbf{Accuracy}.
Figure~\ref{f:diftasks} compares the accuracy of our method with existing approaches. The  split co-teaching approach consistently produced the most accurate substitute QNN across all tasks. 
Our design achieved a 6.5\%$\sim$9.5\% accuracy improvement over the \texttt{base} model and a 0.1\%$\sim$3.7\% enhancement compared to QuantumLeak~\cite{Fu:IJCNN2024}. These results underscore the effectiveness of our approach under varying quantum noise, positioning it as a superior alternative to existing attack methods.

\textbf{Impact of $V_{th}$}.
$V_{th}$ is a critical parameter that determines whether data samples are classified as noise-robust or noise-vulnerable. A lower $V_{th}$ can cause a larger portion of data to be categorized as noise-vulnerable data (i.e., $D_v$), leading to the loss of clean samples during co-teaching. Conversely, a higher $V_{th}$ may retain more noisy samples, potentially disrupting the training process. 
Figure~\ref{f:vth}  illustrates the impact of varying $V_{th}$ on the accuracy of the \texttt{m01} classification task, showing that at $V_{th}=0.15$, our model achieved optimal performance with a peak accuracy of 92.6\% with 72.6\% of the data classified as $D_r$.
For practical implementation, we recommend task-specific profiling to identify the optimal $V_{th}$.

\textbf{Impact of $f_1$ and $f_2$}.
Figures~\ref{f:f1} and ~\ref{f:f2} 
shows results when adjusting the forget rates $f_1$ and $f_2$, respectively.
Increasing the forget rate tends to enhance the proportion of clean data in the training set; however, this reduction in training set size is significant, especially in model extraction attacks where the number of queries is constrained. On the other hand, setting the forget rate too low may fail to adequately remove noisy data, thereby compromising model performance. 
Therefore, both $f_1$ and $f_2$ must be carefully optimized to balance the retention of clean data with the need to maintain a sufficiently large training set. Identifying these optimal values is essential for maximizing performance in model extraction attacks.

\vspace{-0.15in}
\section{Conclusion}
\vspace{-0.1in}
In this work, we introduced a novel attack framework, split co-teaching, designed to address the limitations of existing model extraction methods in practical NISQ environments. By utilizing label variations to partition data based on noise sensitivity and implementing co-teaching strategies, our approach achieves state-of-the-art accuracy in QNN extraction attacks, as validated by results on NISQ computers.

\vspace{-0.2in}
\section*{Acknowledgment}
\vspace{-0.1in}
This work was supported in part by NSF OAC-2417589 and NSF CNS-2143120. 
The authors also acknowledge IBM's support through Quantum Computing Credits.

\clearpage
\bibliographystyle{IEEEbib}
\bibliography{reference}
\end{document}